# Predicted MAX phase Sc$_2$InC: Dynamical stability, vibrational and optical properties

A. Chowdhury[a], M. A. Ali[a], M. M. Hossain[a], M. M. Uddin[a,*], S. H. Naqib[b], A. K. M. A. Islam[b,c,*]

[a]Department of Physics, Chittagong University of Engineering and Technology, Chittagong-4349, Bangladesh
[b]Department of Physics, University of Rajshahi, Rajshahi 6205, Bangladesh
[c]International Islamic University Chittagong,154/A College Road, Chittagong 4203, Bangladesh

**Abstract**

First principles pseudopotential calculations have been performed for the first time to investigate the phonon dispersion, thermodynamic and optical properties including charge density, Fermi surface, Mulliken population analysis, theoretical Vickers hardness of predicted MAX phase Sc$_2$InC. We revisited the structural, elastic and electronic properties of the compound which assessed the reliability of our calculations. The analysis of the elastic constants and the phonon dispersion along with phonon density of states indicates the mechanical stability and dynamical stability of the MAX phase. The Helmholtz free energy, internal energy, entropy specific heat capacity and Debye temperature have also been calculated from the phonon density of states. Mulliken population analysis indicates the existence of prominent covalency in chemical bonding of Sc$_2$InC. The electronic charge density mapping shows a combination of ionic, covalent and metallic bonding in the compound. The Fermi surface is comprised due to the low dispersive Sc 3$d$ and C 2$p$ states from the [ScC] blocks. The phase is expected to be a soft material and easily mechinable due to its low Vicker hardness value. Furthermore, the analysis of various optical properties (such as dielectric function, refractive index, photoconductivity, absorption coefficients, loss function and reflectivity) suggests that the nanolaminate Sc$_2$InC is a promising candidate for optoelectronic devices in the visible and ultraviolet energy regions and as a coating material to avoid solar heating.

## 1. Introduction

The study of MAX phases has become an important sub-branch in materials science and technology due to their interesting properties and various studies since their discovery during 1960s [1-17]. The structure consists of the stacking of n "ceramic" layer(s) inserted by an *A*"metallic" layer [2, 6-8]. A good number of MAX phase compounds have then been synthesized [5, 18-27] and the physical properties investigated [28-41]. For example, the M$_2$AX phases with *M* = (Ti, V, Cr, Nb, Ta, Zr, Hf); *A*= (Al, S, Sn, As, In, Ga) and *X* = (N,C) have been studied both experimentally and theoretically due to their attractive properties as stated earlier [1-5, 42–52]. Due to the continuing efforts by the scientific community existence of further MAX phases has been reported or proposed. As a result the number of synthesized ternary MAX phases has reached over 70 [3, 17] out of a possible 665 viable MAX phases (M$_{n+1}$AX$_n$, n =1–4) [28]. So it is evident that there are still many MAX phases including Sc$_2$InC which till now have not been synthesized. But it may mention that one of the previous studies [52] has shown the chemical stability of this phase by calculating the formation energy.

Previous investigations of the predicted Sc$_2$InC phase, based on first-principles electronic structure calculations, have been reported in the literature [50-52]. Although, the elastic and electronic properties are reported for Sc$_2$InC [50, 51], technologically important physical features like Vickers hardness, Mulliken population analysis, charge density and phonon dispersion as well as thermodynamic and optical properties remain unexplored.

*Corresponding authors.
E-mail address: mohi_cuet@yahoo.com, azi46@ru.ac.bd

Since the phase Sc$_2$InC is predicted and not synthesized yet, the study of dynamical stability is therefore most important for future research and applications. Moreover, understanding the thermodynamic properties can broaden our knowledge on particular behavior of solids under high pressures and temperatures, which are essential for industrial applications [53]. The investigation of optical functions serves as a powerful tool in analyzing the electronic features of solids. Furthermore, studied MAX phases are predicted to be potential coating materials for spacecraft to reduce solar heating significantly [54].

So, we aim to study the mechanical and dynamical stability, the thermodynamic and optical properties of the hypothetical compound Sc$_2$InC including its charge density, Fermi surface, Mulliken population analysis, theoretical Vickers hardness for the first time. In the process we will also revisit some aspects of the elastic and electronic properties in order to assess the reliability of our calculations. The organization of the paper is as follows: Section 2 describes the computational scheme. Results and discussion constitute Section 3. Concluding remarks can be found in Section 4.

## 2. Computational methodology

The first-principles calculations are carried out by employing pseudo-potential plane-waves (PP-PW) approach based on the density functional theory (DFT) [55] as implemented in the CASTEP code [56]. The exchange-correlation potential is treated within the Generalized Gradient Approximation (GGA) method with default Perdew-Burke-Ernzerhof (PBE) [57] and Local Density Approximation (LDA) method with the Ceperley–Alder [58] form. The electron-ion potentials are treated by means of first-principles pseudopotentials within Vanderbilt-type ultrasoft formulation [59] for Sc, In, and C atoms. For k-points sampling integration over the first Brillouin zone, the Monkhorst-Pack scheme [60] is used. The plane wave cut-off energy is taken to be 500 eV. The convergence criteria for structure optimization and energy calculation were set to ultrafine quality with the *k*-point mesh of 9×9×2 for the crystal structure. The tolerance for self-consistent field, energy, maximum force, maximum displacement, and maximum stress are 5.0×10$^{-7}$ eV/atom, 5.0×10$^{-6}$ eV/atom, 0.01 eV/Å, 5.0×10$^{-4}$ Å, and 0.02 GPa, respectively. The Broyden–Fletcher–Goldfarb-Shenno (BFGS) minimization technique is used to optimize the geometry of hexagonal Sc$_2$InC.

## 3. Results and discussion

*3.1. Structural properties*

The structure of ternary layered carbide, Sc$_2$InC, is shown in Fig. 1. This *MAX* phase crystallizes with space group P63/mmc (194) belonging to the hexagonal system. There are eight atoms in each unit cell. The unit cell contains two molecules. The Wyckoff positions of the atoms in $Sc_2InC$ are as follows: C atoms at the positions (0, 0, 0), the $A$ atoms are at (1/3, 2/3, 3/4) and the four Sc atoms are at (1/3, 2/3, $z_M$) [49]. The structure is thus defined by two lattice parameters, $a$ and $c$, and the internal structural parameter, $z_M$. The equilibrium crystal structure of $Sc_2InC$ is first obtained by minimizing the total energy and the optimized values of structural parameters of $Sc_2InC$ compound are given in Table 1. Our results are in accord with the previous theoretical results [50, 52].

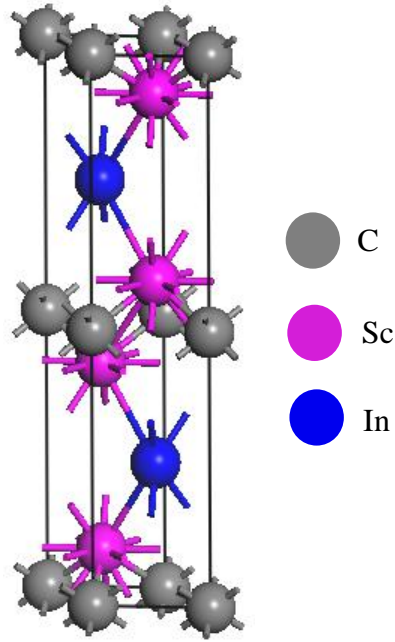

**Fig. 1.** Crystal Structure of $Sc_2InC$.

**Table 1**
Calculated lattice parameters ($a$ and $c$, in Å), hexagonal ratio $c/a$, internal parameter $z_M$, and unit cell volume $V$ (Å$^3$) of $Sc_2InC$ *MAX* compound.

| Phases | $a$ | $c$ | $c/a$ | $z_M$ | $V$ | Ref |
|---|---|---|---|---|---|---|
| $Sc_2InC$ | 3.273 | 15.071 | 4.604 | 0.0804 | 139.814 | This work (GGA) |
| | 3.272 | 16.4516 | 5.028 | | 152.529* | [52] GGA |
| | 3.273 | 15.073 | 4.605 | 0.0804 | 139.865 | This work (LDA) |
| | 3.2752 | 15.082 | 4.605 | 0.0803 | 140.105* | [50] LDA |

*Calculated using $V = 0.866a^2c$

*3.2. Elastic properties*

The calculated five independent elastic constants $C_{ij}$ (since $C_{66} = (C_{11} - C_{12})/2$) and polycrystalline elastic moduli are presented in Table 2 along with available theoretical data [50, 51] which shows that the calculated elastic constants are in reasonable agreement with earlier results. Further our calculated $C_{ij}$ satisfies the Born stability criteria [61]: $C_{11} > 0$, $C_{11}$-

$C_{12}> 0$, $C_{44} > 0$, $(C_{11} +C_{12}) C_{33}- 2C_{13}> 0$. This confirms the mechanical stability of MAX phase Sc$_2$InC.

The bonding nature of Sc$_2$InC can also be investigated from the values of elastic constants. As for example, the small value of $C_{33}$ indicates the compound is more compressible along the c-axis direction compared to along other directions. The value of $C_{11}$ is greater than that of $C_{33}$ indicating the comparatively stronger atomic bonding along the [100] planes between the nearest neighbours than those along the [001] plane. The ratios of $C_{33}$ to $C_{11}$ and $C_{13}$ to $C_{12}$ are not comparable, indicating the dissimilarity of the atomic bondings along the *c*-axis and *a*-axis. The ability to resist the shear distortion in (100) plane is represented by the elastic constant $C_{44}$, whereas the elastic constant $C_{66=}$ $(C_{11} - C_{12})/2$) relates to the resistance to shear in the <110> direction [62]. The result $C_{66} > C_{44}$ demonstrates the compound Sc$_2$InC is able to resist the shear distortion in the <110> direction than in the (100) plane. The shear anisotropy factor *A*, defined by $A = 2C_{44}/(C_{11} - C_{12})$ [63] (Table 2), indicates the anisotropic nature of this compound with the possibility of appearance of microcracks in this material.

**Table 2**

The calculated elastic constants, $C_{ij}$ (GPa), bulk modulus, *B* (GPa), shear modulus, *G* (GPa), Young's modulus, *Y* (GPa), Pugh ratio, *G/B*, and Poisson ratio, ν and anisotropic factor of MAX phase Sc$_2$InC compared with other theoretical results.

| Phases | $C_{11}$ | $C_{12}$ | $C_{13}$ | $C_{33}$ | $C_{44}$ | A | B | G | Y | G/B | ν | Ref |
|---|---|---|---|---|---|---|---|---|---|---|---|---|
| Sc$_2$InC | 209.8 | 52.4 | 40.5 | 193.8 | 68.4 | 0.87 | 97.6 | 74.8 | 178.8 | 0.76 | 0.19 | This work (GGA) |
| | 175 | 59 | 33 | 173 | 41 | 0.72 | 86 | 54 | 135 | 0.63* | 0.24* | [51] GGA |
| | 186 | 43 | 29 | 171 | 44 | 0.62 | 83 | 59 | 173 | 0.73 | 0.21 | This work (LDA) |
| | 206 | 51 | 38 | 191 | 50 | 0.65* | 95 | 65.6 | 160 | 0.69* | 0.22 | [50] LDA |

*Estimated using published data.

The polycrystalline elastic moduli (*B*, *G*, *Y*, and *v*), calculated using Voigt–Reuss–Hill approximations [64] (Table 2), show reasonable agreement with the reported data [50, 51]. The value of *G/B* > 0.5 in both GGA and LDA calculations indicates that the phase is brittle in nature following Pugh's criterion [65]. A material can be separated from the ductile to brittle in terms of Poisson's ratio (*v*) [66]. This rule proposes *v* ~ 0.26 as the border line that separates the brittle and ductile materials. Our calculated value of *v* (0.19 for GGA, 0.21 for LDA) lies in the lower side of the critical showing the brittleness of the phase. Further *v* is typically small (0.1) for covalent materials and *v* ~ 0.25 for ionic materials [67]. The obtained value of *v* (0.19 for GGA, 0.21 for LDA) for the Sc$_2$InC shows that it is in the border line of covalent/ionic materials.

*3.3. Electronic properties*

*3.3.1 Band structure and DOS*

The calculated energy band structure of $Sc_2InC$ with *k* points in the first Brillouin zone using equilibrium lattice parameters is shown in Fig. 2a. The compound exhibits metallic nature since its conduction bands (red colours) cross the $E_F$ line and overlap noticeably with the valence bands. In addition, no band gap is found. The near-Fermi bands show a complicated 'mixed' character, combining the quasi-flat bands with a series of high-dispersive bands intersecting the $E_F$. The detailed features of band structure can be explained with the calculated total and partial density of states.

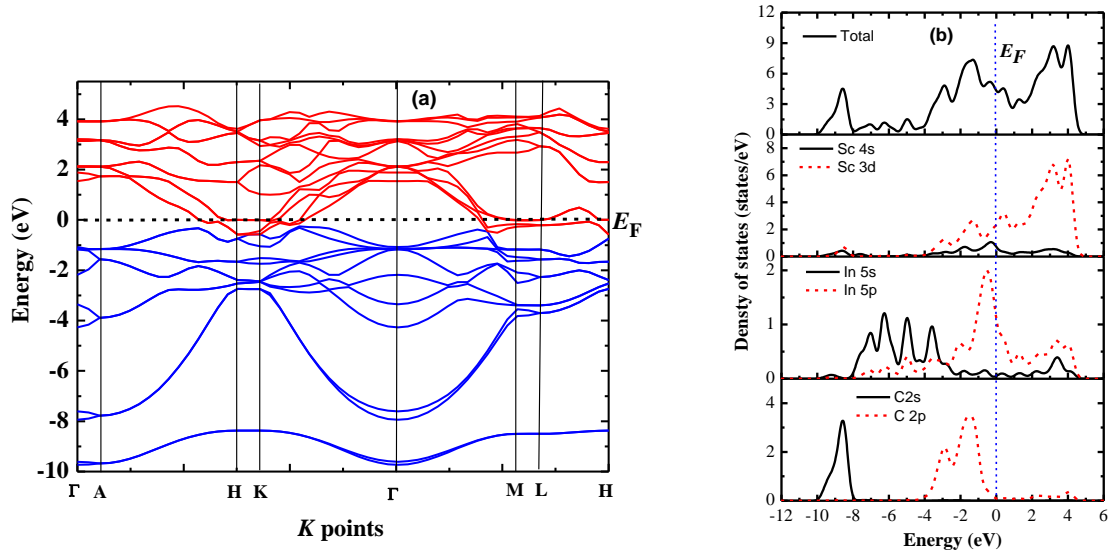

**Fig. 2.** (a) Electronic band structure and (b) the total and partial density of states (DOS) of $Sc_2InC$ compound.

The valence band can be resolved into three sub-bands. The lowest energy sub-band (-8 eV to -10 eV) is derived mainly from C-2*s* states. The middle sub-band of broad nature is formed primarily from the contributions due to C-2*p* and In-5*s* orbitals. The top sub-band crossing the Fermi level is derived mainly from Sc-3*d* and In-5*p* electronic states.

The calculated total and partial densities of states are shown in Fig. 2(b), where the vertical broken line denotes the Fermi level, $E_F$. The total energy scale is chosen from -12 eV to 6 eV. The value of DOS at $E_F$ is found to be 4.4 states per eV. From the analysis of the density of states we observe that Sc-3*d* electrons are mainly contributing to the DOS at the Fermi level, and should be involved in the conduction properties. The contributions from In-s and p states are also noticeable but an order of magnitude smaller than that of Sc-3*d* states. C-2*s* and -2*p* states do not contribute to the DOS at the Fermi level and therefore is not involved in the charge conduction. The valence band can be resolved into three sub-bands. The middle sub-band of broad nature is formed primarily from the contributions due to C-2*p* and In-5*s* orbitals. The top sub-band crossing the Fermi level is derived mainly from Sc-3*d* and In-5*p* electronic states.

*3.3.2 Mulliken atomic and bond overlap populations*

The effective valence charge (*EVC*) and bond overlap population (*BOP*) analysis have been made using the Mulliken atomic population (*MAP*) for which formalism can be found elsewhere [68]. The *MAP* and the *EVC* effectively provides information regarding the nature of chemical bonding. The difference between the formal ionic charge and the Mulliken charge on the anion species is called the *EVC*. It is used to predict the strength of a bond either as covalent or ionic and so on. The zero value of *EVC* is for an ideal ionic bond while departure from zero measures the degree of covalency. The calculated effective valence is presented in Table 3 which indicated the existence of prominent covalency in chemical bonding inside the $Sc_2InC$ compound.

**Table 3**

Mulliken atomic and bond overlap population of the $Sc_2InC$ compound.

| | Mulliken atomic population | | | | | | Mulliken bond overlap population | | | |
|---|---|---|---|---|---|---|---|---|---|---|
| Atoms | $s$ | $p$ | $d$ | Total | Charge (e) | EVC (e) | Bond | Bond number $n^\mu$ | Bond length $d^\mu$ (Å) | Bond overlap population $P^\mu$ |
| C | 1.49 | 3.34 | 0.00 | 4.83 | -0.83 | --- | C-Sc | 4 | 2.24466 | 1.13 |
| Sc | 2.24 | 6.63 | 1.58 | 0.54 | 0.54 | 2.46 | Sc-Sc | 2 | 3.07237 | -0.38 |
| | | | | | | | Sc-In | 4 | 3.17959 | 0.11 |
| In | 1.35 | 1.93 | 9.98 | 13.25 | -0.25 | 3.25 | C-In | 4 | 4.21555 | -0.12 |

The value of the *BOP* indicates the bonding and antibonding states in the compound by the positive or negative value, respectively. The value close to zero indicates no significant interaction between the electronic populations of two atoms. The higher the value of *BOP* represents higher level of covalency. It seems that the strong covalent bond between C-Sc exists in the $Sc_2InC$ compound. It is found from the Mulliken population analysis that the nature of $Sc_2InC$ is metallic with some covalent and ionic.

*3.3.3 Charge Density and Fermi surface*

In order to unfold the nature of chemical bonding in the $Sc_2InC$ compound electron charge density distribution mapping in contour form (in the units of $e/Å^3$) along (101) crystallographic plane has been calculated and shown in Fig. 3(a). The colors red and blue in the adjacent scale to the map indicate the high and low value of electronic charge density, respectively.

It is seen from the charge density mapping that there are strong accumulation of charges in the Sc and C regions, but charges are depleted in the In regions. The bonds Sc-C can be considered as strong covalent bond, which arises from the strong hybridization of Sc-3$d$ and C-2$p$ electrons at the Fermi level. It is observed clearly that C-In bonds are ionic in nature. The bond Sc and In states lead to the formation of weak covalent bonds. Furthermore, metallic type bonds are expected to exist in the blocks Sc-Sc atoms. The strong metallic bonding among Sc atoms results in the good conductivity of

Sc$_2$InC which is consistent with the DOS of Sc$_2$InC. A highly anisotropic combination of chemical bonds ionic, covalent and metallic interactions exists in the Sc$_2$InC compound.

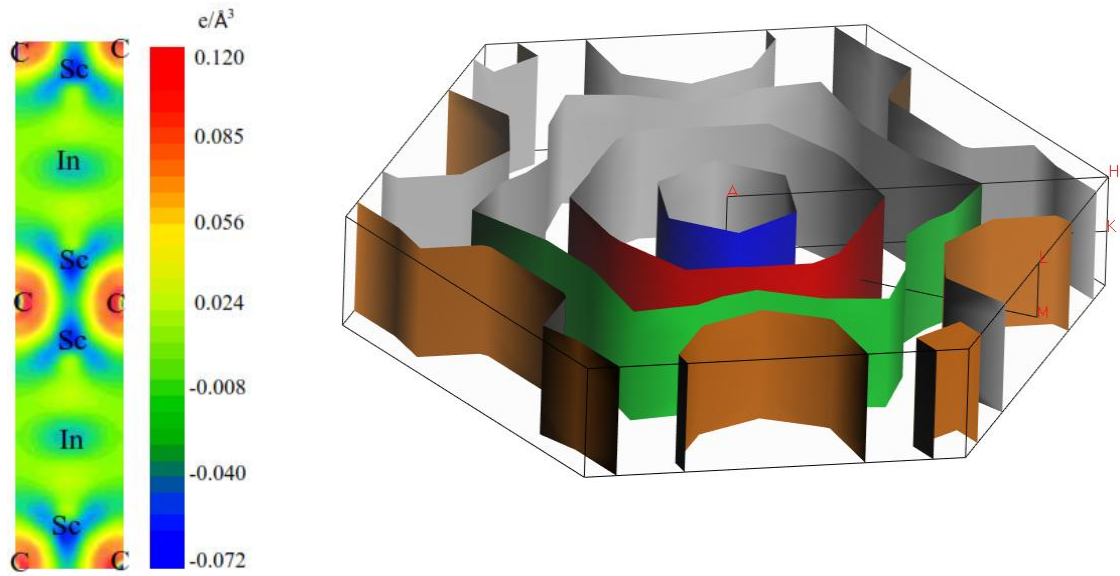

**Fig. 3.** (a) Charge density (left) and (b) Fermi surface (right) of Sc$_2$InC compound.

The Fermi surface topology of the Sc$_2$InC has been investigated in the equilibrium structure at zero pressure (Fig. 3b). Electron and hole-like sheets are present in the calculated Fermi surface. The central sheet (inner blue) is of cylindrical shape with hexagonal cross section and centered along the Γ-A direction of the Brillouin zone. Second sheet (red colour) is also cylindrical in shape, and has a larger cross section. The third sheet is little bit complicated and it is expanded along Γ–K direction and shrink along Γ–M directions. The fourth sheet (outer most) consists of six up curved ribbon type tubes in each corner. The Fermi surface of Sc$_2$InC is due to the low dispersive Sc 3$d$ and C 2p states from the [ScC] blocks, which can also be confirmed from DOS in Fig. 2. All these sheets are separated from each other.

*3.4 Calculation of Vickers Hardness*

The relevant formula for the hardness expressed via Mulliken bond populations is given as [69, 70]:

$$H_V = \left[ \prod^{\mu} \left\{ 740(P^{\mu} - P^{\mu'})(v_b^{\mu})^{-5/3} \right\}^{n^{\mu}} \right]^{1/\sum n^{\mu}},$$

Where $P^{\mu}$ is the Mulliken population of the $\mu$-type bond, $P^{\mu'} = n_{free}/V$ is the metallic population, and $v_b^{\mu}$ is the bond volume of $\mu$-type bond. The obtained results from the calculations are given in Table 4. The Vickers hardness value of 2.4 GPa is comparable to those obtained in other MAX phase compounds such as Mo$_2$GaC with $H_v$ = 2.69 GPa [37], indicating the soft nature of Sc$_2$InC.

**Table 4**
Muliken bond overlap population of $\mu$-type bond $P^\mu$, bondlength $d^\mu$, metallic population $P^{\mu'}$, bond volume $v_b^\mu$, Vickers hardness of $\mu$-type bond $H_V^\mu$ and $H_V$ of $Sc_2InC$.

| Phase | Bond | $d^\mu$ | $P^\mu$ | $P^{\mu'}$ | $v_b^\mu$ | $H_V^\mu$ | $H_V$ |
|---|---|---|---|---|---|---|---|
| $Sc_2InC$ | Sc-C | 2.24474 | 1.13 | 0.0258 | 9.107 | 20.59 | 2.4 |
|  | Sc-In | 3.17916 | 0.11 |  | 25.87 | 0.2753 |  |

*3.5 Phonon dispersion – Dynamical stability*

The phonon dispersion curve (PDC) of a material provides information regarding structural stability and vibrational contribution in the thermodynamic properties such as thermal expansion, Helmholtz free energy, and heat capacity [71, 72]. The PDC and phonon density of states (PHDOS) of the $Sc_2InC$ have been calculated along the high symmetry direction of the crystal Brillouin zone (BZ) using the DFPT (Density Functional Perturbation Theory) linear-response method [73] and shown in Fig. 4. The optical branch is situated at the top in the PDC shown in Fig. 4 (a) that is responsible of the most optical behaviours of the compound. The value of corresponding frequencies for transverse optical (TO) and longitudinal optical (LO) phonon modes at the zone center ($\Gamma$) are found to be 12.8 and 18.7 THz, respectively.

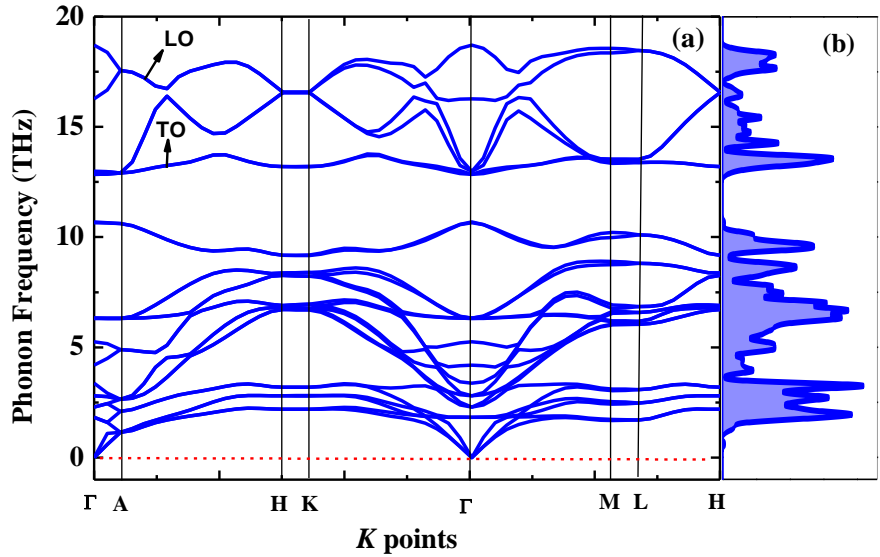

**Fig. 4.** (a) Phonon dispersion curve (PDC) and (b) phonon density of states (PHDOS) of $Sc_2InC$. The dashed line (red) is at zero phonon frequency. No imaginary (negative) frequency exists.

The PHDOS is presented in Fig. 4 (b) side by side with the PDC for better understanding of the bands by comparing corresponding peaks. The flatness of the bands for TO shown in Fig. 4 (a) gives a pronounced peak [Fig. 4(b)] while non-flat bands for LO in Fig. 4(a) results in weak peaks [Fig. 4(b)]. It is noteworthy that a clear gap exists between the acoustic and optical branches. In addition, the top of the LO and bottom of the TO modes are located at the Γ point and the separation between these is 5.9 THz. A compound is said to be dynamically stable (unstable) if the phonon frequencies for all wave vectors are positive (negative). The phonon dispersion curve (over the whole BZ) does not show imaginary (negative in the frequency scale) phonon frequency indicating that $Sc_2InC$ phase is dynamically stable. Furthermore, the elastic constants of $Sc_2InC$ satisfy the Born conditions for mechanical stability.

*3.6 Thermodynamic properties*

The thermodynamical potential functions such as Helmholtz free energy *F*, internal energy *E*, entropy *S*, specific heat $C_v$ and Debye Temperature $\Theta_D$ of $Sc_2InC$ are calculated at zero pressure using quasi-harmonic approximation [32, 74].

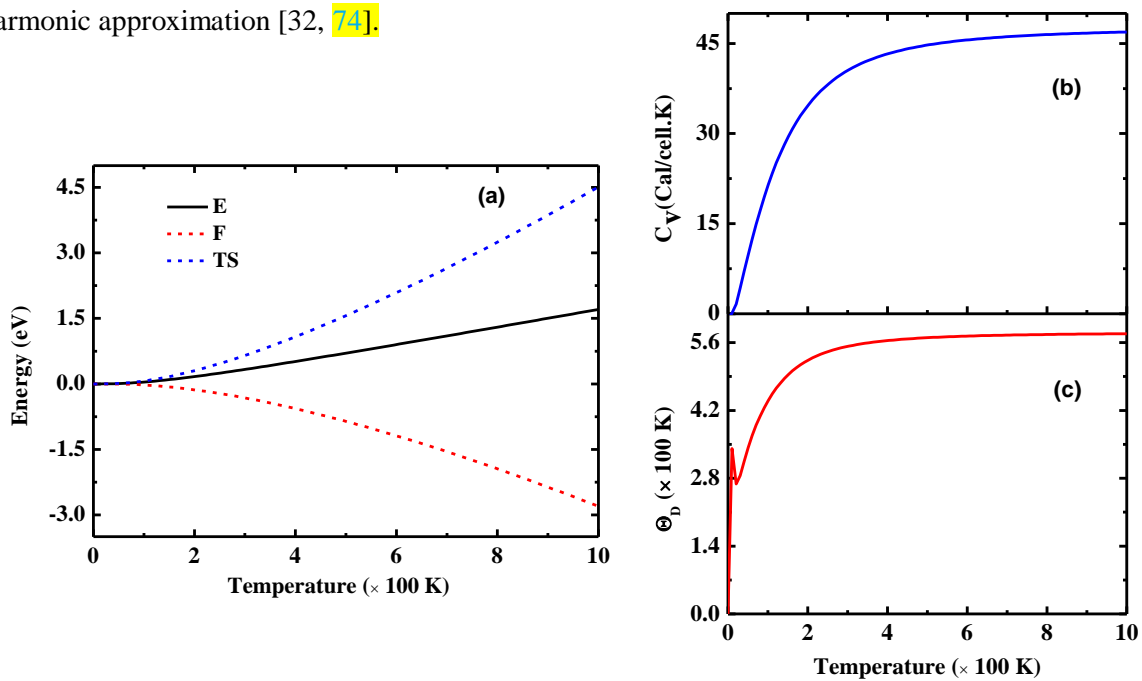

**Fig. 5.** Temperature dependence of the thermodynamical potential functions of $Sc_2InC$.

The calculated *F*, *E*, *S*, $C_v$ and $\Theta_D$ of the $Sc_2InC$ compound are displayed in Fig. 5(a, b and c) in the temperature range from 0 to1000 K in which the harmonic model is assumed to be valid. Fig. 5(a) shows the temperature dependence of Helmholtz free energy (*F*) where it is found to decrease gradually with the increase of temperature. The decreasing trend of free energy is very common and it becomes more negative during the course of any natural process. The free energy is defined as the difference between internal energy of a system and the amount of energy that cannot be used to perform work. This unusable energy is expressed as the product of entropy of a system and the absolute temperature of the system. In order to compare Helmholtz free energy and internal energy we present the entropy as TS. Here S is the lattice entropy resulting from lattice vibration and this entropy rises with increasing temperature. The calculated free energy is also found to be decreased with increasing temperature as thermal disturbance adds to disorder. Fig. 5(a) also show the variation of internal energy (*E*) with temperature in which an increasing trend with temperature is observed as expected for solids.

Fig. 5 (b) shows that the specific heat $C_V$ of the Sc$_2$InC compound. As temperature increases phonon thermal softening occurs and as a result the heat capacities increase with increasing temperature. In the low temperature limit, $C_V$ of Sc$_2$InC follows the Debye model which is proportional to $T^3$, as expected [75]. The curve follows Dulong-Petit law at high temperatures [76]. Above 400 K, $C_V$ increases slowly with temperature and gradually approaches the Dulong–Petit limit.

Fig. 5(c) also displays the temperature dependence of Debye temperature $\Theta_D$ as calculated from PHDOS. It is seen that $\Theta_D$ increases with increasing temperature indicating the change of the vibration frequency of particles under temperature effects. It is also related to the bonding strength. The weaker bonds in solids, the lower the value of $\Theta_D$, the heat capacity reach the classical Dulong–Petit value at a lower temperature. $\Theta_D$ is also estimated through the calculation of average sound velocity using the formalism described elsewhere [68]. The calculated $\Theta_D$ along with sound velocities $v_l$, $v_t$, and $v_m$ are presented in Table 5 from where the value is comapered with the available reported result [50].

**Table 5.**
Calculated density ($\rho$ in gm/cm$^3$), longitudinal, transverse and average sound velocities ($v_l$, $v_t$, and $v_m$ in km/s) and Debye temperature ($\theta_D$ in K) of Sc$_2$InC.

| Compound | $\rho$ | $v_l$ | $v_t$ | $v_m$ | $\Theta_D$ | Ref. |
|---|---|---|---|---|---|---|
| Sc$_2$InC | 5.15 | 6.20 | 3.80 | 4.21 | 483 | This work |
|  | 5.14 | 5.96 | 3.57 | 3.95 | 453 | [50] |

*3.7 Optical properties*

We have calculated the optical properties of Sc$_2$InC for photon energies up to 20 eV for polarization vectors [100] and [001] and these displayed in Fig. 6 (a-h). We have used a 0.5 eV Gaussian smearing for all calculations. This smears out the Fermi level, so that k-points will be more effective on the Fermi surface. The curves for two polarization directions are similar in nature but not identical, however the peak positions are different and distinguishable with height as well. Electron transitions takes place from occupied states below the $E_F$ to unoccupied states above the $E_F$, when light of sufficient energy shines onto a material. The low energy infrared part of the spectra in metal and metal-like systems is affected mainly by the intra-band contribution to the optical properties of the materials. A semi-empirical Drude term is employed to calculate the dielectric function. A Drude term with unscreened plasma frequency 3 eV and damping 0.05 eV has been used [77].

The optical properties of Sc$_2$InC are determined by the frequency-dependent dielectric function $\varepsilon(\omega)=\varepsilon_1(\omega)+i\varepsilon_2(\omega)$. The imaginary part $\varepsilon_2(\omega)$ of the dielectric function $\varepsilon(\omega)$ is calculated from the momentum matrix elements between the occupied and unoccupied electronic states and given by [78]

$$\varepsilon_2(\omega) = \frac{2e^2\pi}{\Omega\varepsilon_0} \sum_{k,v,c} |\psi_k^c|\mathbf{u}\cdot\mathbf{r}|\psi_k^v|^2 \delta(E_k^c - E_k^v - E)$$

Here $\mathbf{u}$ is the vector defining the polarization of the incident electric field, $\omega$ is the light frequency, $e$ is the electronic charge and $\psi_k^c$ and $\psi_k^v$ are the conduction and valence band wave functions at $k$, respectively. The real part $\varepsilon_1(\omega)$ of the dielectric function $\varepsilon(\omega)$ can be derived from the imaginary part $\varepsilon_2(\omega)$ using the Kramers–Kronig dispersion equation. All other energy dependent optical constants - the absorption spectrum, refractive index, extinction coefficient, energy-loss spectrum, and reflectivity are as those given by Eqs. 49–54 in Ref. [78].

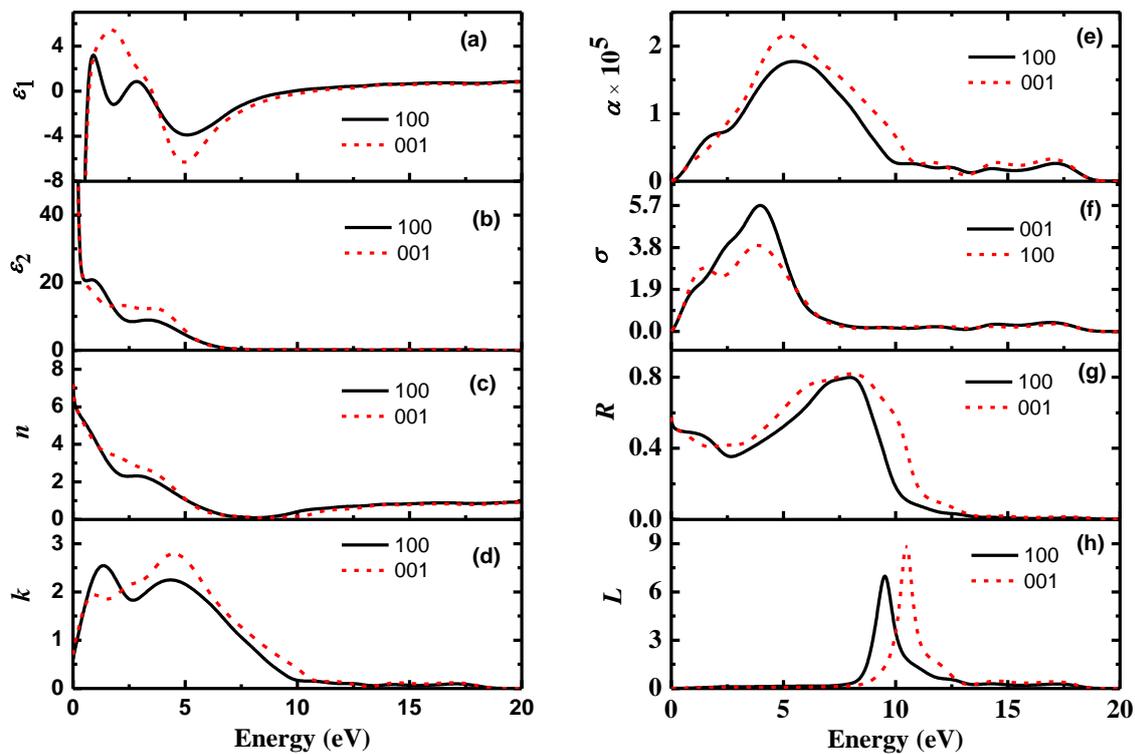

**Fig. 6.** Energy dependent (a) real part of dielectric function, (b) imaginary part of dielectric function, (c) refractive index, (d) extinction coefficient, (e) absorption coefficient, (f) photo conductivity, (g) reflectivity, (d) loss function of Sc$_2$InC compound for two polarization directions.

The real part $\varepsilon_1(\omega)$ and imaginary part of $\varepsilon_2(\omega)$ of the dielectric function of Sc$_2$InC are shown in Figs. 6 (a, b). The peaks in $\varepsilon_2(\omega)$ are associated with the electron excitations. The maximum sharp peak for $\varepsilon_2(\omega)$ was obtained at ~ 0.1 eV, which represents the maximum absorption behaviour of this material. The electronic band structure of Sc$_2$InC was observed as metallic in nature, which is also identified from Fig. 6 a. There is only one broad prominent peak at 3.7 eV (Fig. 6b). The large negative value of $\varepsilon_1$ is also observed in Fig. 6a, indicates a clear sign of Drude-like behavior as seen in metals. $\varepsilon_1(\omega)$ became zero at around 10 eV, which corresponds to the energy at which the absorption coefficients vanishes (Fig. 6e), reflectivity exhibits a sharp drop (Fig. 6g) and the energy loss function (Fig. 6h) shows a first peak. Moreover, the Fig. 6(a, b) exhibit that the real part of the dielectric function goes through zero from below and the imaginary part of the dielectric function approaches zero from above which indicates that studied compound is metallic in nature.

The refractive index, $n$ and extinction coefficient, $k$ of complex refractive index for $Sc_2InC$ compound are displayed in Figs. 6 c and d, respectively. The static refractive index $n$ (0) is found to be 7.2. The nature of the variation of the refractive index and extinction coefficient roughly follows the imaginary and real part of complex dielectric function, respectively.

The absorption coefficients of $Sc_2InC$ compound is shown in Fig. 6 (e) which begins at 0 eV due to their metallic nature. A strong absorption coefficient is observed in the UV region, however it is weak in the IR region but continuously increases toward the UV region, and reaches a maximum value at 5.47 eV. These results indicate that the $Sc_2InC$ compound is promising for absorbing materials in the UV region. The absorption of photons is increased in the material with high absorption coefficient and there by exciting the electrons from the valence band to the conduction band. Therefore, this stimulating properties makes $Sc_2InC$ compound is very important for optical and optoelectronic devices in the visible and ultraviolet energy regions.

The band structures of $Sc_2InC$ shows no band gap indicating the photoconductivity starts at zero photon energy (Fig. 6f). A very good metallic nature of the $Sc_2InC$ compound is confirmed by this type of photoconductivity. The photoconductivity data are in good agreement with the bands structure result as well.

The reflectivity spectra as a function of photon energy of the $Sc_2InC$ compound are shown in Fig. 6 (g). The reflectivity curve shows that it starts with a value of 0.56, decreases and then rises again to a maximum value of 0.79 at 8.15 eV. It is noteworthy that reflectivity is always above 44%. According to Li *et al*. [79] a compound will be capable of reducing solar heating if it has reflectivity ~ 44% in the visible light region. Therefore, the $Sc_2InC$ compound is also a promising candidate for the practical usage as a coating material to avoid solar heating.

The energy loss function (*L*) shown in Fig. 6(h) represents the energy loss of a fast electron passing though the material. The loss function shows a peak, which corresponds to the so-called bulk plasma frequency $\omega_P$. The peak in energy-loss function ascends, however the $\varepsilon_1(\omega)$ goes through zero from below while $\varepsilon_2(\omega)$ goes through zero from above. In the energy-loss spectrum, we see that the effective plasma frequency $\omega_P$ of the compound is equal to ~ 9.5 eV. The material becomes transparent, when the frequency of the incident light is higher than that of plasma frequency. Moreover, the peaks of loss function as shown in Fig. 6 (h) correspond to the trailing edges in the reflection spectra.

## 4. Conclusions

The first-principle pseudopotential calculations been employed to study dynamic stability of hypothetical MAX phase $Sc_2InC$, in addition to investigating thermodynamic and optical properties, electronic charge density, Fermi surface, Mulliken bond overlap population and Vickers hardness. The evaluated lattice parameters, elastic constants $C_{ij}$ including the polycrystalline elastic constants show a good consistency with the available results. The elastic constants obey the traditional mechanical stability conditions. The brittleness of the compound has been shown to be similar as expected for other MAX phases. The electronic band structure, DOS, electron charge density mapping, Fermi surface, and Mulliken atomic and bond overlap populations have been calculated and analysed. The results show that $Sc_2InC$ is metallic in nature where the contribution from Sc-3$d$ states dominates the electronic conductivity at the Fermi level. The results also reveal that the compound possesses intra-atomic bonding with a mixture of ionic, covalent and metallic interactions. The obtained Vicker hardness (2.4 GPa) indicates the soft nature of $Sc_2InC$.

The calculated phonon dispersive curve confirms the dynamical stability of $Sc_2InC$ MAX phase. As temperature increases phonon thermal softening occurs and as a result the heat capacities increase with increasing temperature. The value Debye temperature $\Theta_D$ is found to be 483 K. The analysis of the calculated optical properties reveals several interesting properties of the phase. The compound is expected to be a promising candidate for optoelectronic devices in the visible and ultraviolet energy regions and as a coating material to avoid solar heating.